\documentclass[pra,twocolumn,showpacs,superscriptaddress,floatfix,amsmath,nofootinbib,amssymb]{revtex4} 

\usepackage[english]{babel}
\usepackage{graphicx}
\usepackage{dcolumn}
\usepackage{bm}
\usepackage{verbatim}
\usepackage{color}
\usepackage{soul}

\newcommand{\ket}[1]{\left| {#1} \right\rangle}
\newcommand{\bra}[1]{\left\langle {#1} \right|}

\newcommand{\proj}[2]{\left| {#1} \right\rangle\!\left\langle {#2} \right|}

\usepackage{color,xspace}

\def\slashchar#1{\setbox0=\hbox{$#1$} 
\dimen0=\wd0 
\setbox1=\hbox{/} \dimen1=\wd1 
\ifdim\dimen0>\dimen1 
\rlap{\hbox to \dimen0{\hfil/\hfil}} 
#1 
\else 
\rlap{\hbox to \dimen1{\hfil$#1$\hfil}} 
/ 
\fi}

\begin{document}

\title{Particle and anti-particle bosonic entanglement in non-inertial frames}

\author{David Edward Bruschi}
\address{School of Mathematical Sciences, University of Nottingham, Nottingham NG7 2RD, United Kingdom}
\author{Andrzej Dragan}
\address{School of Mathematical Sciences, University of Nottingham, Nottingham NG7 2RD, United Kingdom}
\address{Institute of Theoretical Physics, University of Warsaw, Ho\.{z}a 69, 00-049 Warsaw, Poland}
\author{Ivette Fuentes\footnote{Previously known as Fuentes-Guridi and Fuentes-Schuller.}}
\author{Jorma Louko}
\address{School of Mathematical Sciences, University of Nottingham, Nottingham NG7 2RD, United Kingdom}

\begin{abstract}
We analyse the entanglement tradeoff between particle and anti-particle modes of a charged bosonic field between inertial and uniformly accelerated observers. In contrast with previous results for fermionic fields, we find that the entanglement redistribution between particle and antiparticle modes does not prevent the entanglement from vanishing in the infinite acceleration limit.

\end{abstract}

\maketitle

\section{Introduction}
One of the main aims in the emerging field of Relativistic Quantum Information is to understand how entanglement depends on the motion of an observer. It has been shown that the amount of entanglement present in a state of free modes of a relativistic quantum field as seen by inertial observers, such as Alice and Bob, is degraded when the same state is analyzed by a uniformly accelerated observer Rob \cite{Alsingtelep,Alicefalls,AlsingSchul,Adeschul,AlsingMcmhMil,Edu2,schacross,Hu,Edu6,Friis}. The vacuum state as seen by Alice is a highly populated thermal state for Rob. General states for inertial observers, including entangled states, inherit the noise introduced by the vacuum. Therefore, the states as seen by an accelerated observer would not be thermal but more mixed. The amount of noise would increase with the acceleration while degrading correlations. In this sense, the Unruh effect degrades entanglement\cite{Alsingtelep,Alicefalls,AlsingSchul,Adeschul,AlsingMcmhMil,Edu2,schacross,Hu,Edu6,Friis}. 

The temperature of the Minkowski vacuum diverges in the infinite acceleration limit. As a consequence, it was found that correlations present in maximally entangled states of uncharged bosonic field modes vanish in the infinite acceleration limit. Surprisingly, for fermionic field modes, entanglement remains finite in the limit of infinite acceleration (for example, see \cite{Edu6}). The reasons for this striking difference are not yet understood. In order to address this issue, nonlocal correlations between fermionic particle and antiparticle degrees of freedom have also been taken into account \cite{EDUIVY}. There the authors considered initially maximally entangled states and three different bipartitions: the first where Rob could not distinguish between particle and antiparticles and two where he could analyze separately particles and antiparticles. They found that the survival of entanglement in the infinite acceleration in the first bipartition could be accounted for by considering the redistribution of entanglement between particle and antiparticle bipartitions. 
While \cite{EDUIVY} did improve the understanding of the behavior of fermionic entanglement as described by different observers, the behavior could not be directly compared with that for bosons, as previous work on bosons has focused on real scalar fields in which there is no distinction between particles and antiparticles.

In this work we introduce charged bosonic fields. Alice and Bob will analyze a one parameter family of maximally entangled states of Unruh modes. Bob and uniformly accelerated Rob will not agree on the particle content of each of these states. We consider the same bipartitions and analyze the bosonic analogue of the states studied in \cite{EDUIVY}. We study the entanglement tradeoff between such bipartitions and how the entanglement is degraded as a function of the Rob's proper acceleration.
In spite of the presence of antiparticles, we find that mode entanglement always vanishes in the infinite acceleration limit. The redistribution of entanglement between particles and antiparticles observed in the fermionic case \cite{EDUIVY} does not occur for charged bosons. This supports the conjecture that the main differences in the behavior of entanglement in the bosonic and fermionic case are due to Fermi-Dirac versus Bose-Einstein statistics \cite{Edu5}.

This paper is organized as follows:  in section \ref{sec1} we introduce transformations between Minkowski, Unruh and Rindler modes for charged bosonic fields. This section extends results from \cite{EDUIVY,Jorma} and by including anti-particle bosonic modes. In section \ref{sec2}, we analyse the entanglement transfer between the particle and antiparticle sectors in different families of maximally entangled states when one of the observers is uniformly accelerated. Finally, conclusions and discussions are presented in section \ref{conclusions}.\\
Our conventions are the following: $\hbar=c=1$ and the signature of the metric is $(-,+)$. In the following, $M$ and $R$ will be short notation for Minkowski and Rindler.

\section{Charged bosonic field states for uniformly accelerated observers}\label{sec1}

We consider a free charged scalar field $\Phi$ in $1+1$ Minkowski spacetime with metric $g_{\mu\nu}$. For the purposes of this work we consider a massless field without loss of generality.  $\Phi$ obeys the Klein Gordon equation $\Box\Phi=0$  where $
\Box=(\sqrt{-g})^{-1}\partial^{\mu}\sqrt{-g}\partial_{\mu}$ and $g=\text{det}(g_{\mu\nu})$. Inertial observers Alice and Bob will naturally use Minkowski coordinates $(t,x)$ to describe the field. In these coordinates the line element takes the form $ds^2=g_{\mu\nu}dx^{\mu}dx^{\nu}=-dt^2+dx^2$ and $\Box=\partial^{\mu}\partial_{\mu}$. Particular solutions $u^{\pm}_{\omega,\text{M}}(t,x)$ to these field equations take the standard form which can be found in \cite{BIRRELLANDDAVIES}.  For $1+1$ massless scalar, right movers $u^{\pm}_{\omega,\text{M}}(t-x)$ and left movers $u^{\pm}_{\omega,\text{M}}(t+x)$ decouple and we will choose to deal with right movers without loss of generality. These modes satisfy the eigenvalue equation 
\begin{align}
i\partial_t u^{\pm}_{\omega,\text{M}}(t,x)=&\pm\omega u^{\pm}_{\omega,\text{M}}(t,x),
\end{align}
where $\omega>0$ is the frequency with respect to $\partial_t$. We therefore interpret $u^+_{\omega,\text{M}}$ as Minkowski particles and $u^{-}_{\omega,\text{M}}$ as antiparticles. We define the inner product between two modes in any coordinate chart as
\begin{equation}
(u^{\sigma}_{\omega},u^{\sigma'}_{\omega'})=i\int_{\Sigma} d\Sigma^{\mu}\, u_{\omega}^{\sigma*} \overset{\leftrightarrow}{\partial_{\mu}} u^{\sigma'}_{\omega'},\label{inner:product}
\end{equation}
where in Minkowski coordinates we can choose $\Sigma$ as the  $t=0$ hypersurface and $\sigma,\sigma'=\pm$. Modes are (delta) normalized through \eqref{inner:product}: $(u^{\pm}_{\omega},u^{\pm}_{\omega'})=\delta(\omega-\omega')$ and the mixed products vanish.

The field $\Phi$ can then be expanded in terms of particle and antiparticle solutions 
\begin{equation}
\Phi=\int d\omega\left(c_{\omega,\text{M}}\, u^+_{\omega,\text{M}}+ d_{\omega,\text{M}}^\dagger \ u^{-}_{\omega,\text{M}}\right),\label{minkowski:field}
\end{equation}
where $c_{\omega,\text{M}},d_{\omega,\text{M}},c_{\omega,\text{M}}^{\dagger},d_{\omega,\text{M}}^{\dagger}$ are the annihilation and creation operators of particles and antiparticles respectively. They define the Minkowski vacuum $\ket{0}_M$ through the condition $c_{\omega,\text{M}}\ket{0}_M=d_{\omega,\text{M}}\ket{0}_M=0$ and obey the standard commutation relations
\begin{equation}
\left[c_{\omega,\text{M}},c_{\omega',\text{M}}^{\dagger}\right]=\left[d_{\omega,\text{M}},d_{\omega',\text{M}}^{\dagger}\right]=\delta(\omega-\omega'),
\end{equation}
while all other commutators vanish. 

A uniformly accelerated observer Rob follows trajectories which are naturally described by a set of Rindler coordinates $(\eta,\xi)$ and are parametrized by $\xi=\text{const}$. 
The relation between Minkowski and Rindler coordinates is 
\begin{align}
\label{rindler:transformation}
t=\xi\sinh \eta\nonumber\\
x=\xi\cosh\eta,
\end{align}
where $\eta$ is the dimensionless time parameter which increases towards the future, $\xi>0$, $|t|<x$, and the line element reads $ds^2=-\xi^2d\eta^2+d\xi^2$. The part of Minkowski spacetime covered by $(\eta,\xi)$ is called right Rindler Wedge or region $I$. Analogously, one can define a region $II$ by 
\begin{align}
\label{rindler:transformation:two}
t=\xi\sinh \eta\nonumber\\
x=\xi\cosh\eta,
\end{align}
where  $|t|<-x$, $\xi<0$ and $t\rightarrow+\infty$ for $\eta\rightarrow-\infty$. In both regions Rob, moving on a uniformly accelerated world line at $\xi=\text{const}$, would perceive constant proper acceleration of magnitude $A=1/|\xi|$.

One wishes to find the equivalent expansion for Rob of \eqref{minkowski:field}. In his coordinates the operator $\Box$ takes a different form than in the inertial case. Solutions in region $I$ are $u^{\pm}_{\Omega,\text{I}}(\eta,\xi)$ can be found analytically \cite{Jorma}. $\Omega>0$ is a dimensionless frequency and these solutions satisfy
\begin{align}
i\partial_\eta u^{\pm}_{\Omega,\text{I}}(\eta,\xi)=&\pm\Omega u_{\Omega,\text{I}}(\eta,\xi).
\end{align}
One can (delta) normalize such modes through \eqref{inner:product}, where $\Sigma$ will be the $\eta=0$ hyper surface, and obtain $(u^{\pm}_{\Omega,\text{I}},u^{\pm}_{\Omega',\text{I}})=\delta(\Omega-\Omega')$. Analogously, one can follow the same procedure for region $II$.
The field $\Phi$ can then be expanded as
\begin{equation}
\Phi\!=\int d\Omega\left(c_{\Omega,\text{I}} u^+_{\Omega,\text{I}}+ d_{\Omega,\text{I}}^\dagger u^{-}_{\Omega,\text{I}}+c_{\Omega,\text{II}} u^+_{\Omega,\text{II}}+ d_{\Omega,\text{II}}^\dagger  u^{-}_{\Omega,\text{II}}\!\right),
\end{equation}
where $c_{\Omega,\Delta}$ $d_{\Omega,\Delta}$ are the particle and antiparticle operators of regions $\Delta=I,II$. They obey the standard commutation relations
\begin{equation}
\left[c_{\Omega,\Delta},c_{\Omega',\Delta}^{\dagger}\right]=\left[d_{\Omega,\Delta},d_{\Omega',\Delta}^{\dagger}\right]=\delta(\Omega-\Omega').
\end{equation}
Mixed commutators and commutators of operators in different regions vanish and these operators annihilate the Rindler vacuum  $c_{\Omega,\Delta}\ket{0}_R=d_{\Omega,\Delta}\ket{0}_R=0$ $\forall\Omega>0$ and $\Delta=I,II$.

Equation \eqref{inner:product} can be used to compute the Bogoliubov transformations between Minkowski operators and Rindler operators. For example, $c_{\Omega,I}=(\Phi,u^+_{\Omega,\text{I}})$ and one can expand $\Phi$ in the Minkowski basis to find the relation between $c_{\Omega,I}$ and $c_{\omega,\text{M}},d_{\omega,\text{M}},c_{\omega,\text{M}}^{\dagger},d_{\omega,\text{M}}^{\dagger}$. For the purpose of understanding how entanglement is affected by the state of motion of one observer, it is much more convenient to employ Unruh modes.

It is well known that the Unruh basis provides an intermediate step between Minkowski and Rindler modes and allows for analytical Bogoliubov transformation between Unruh operators and Rindler operators.
Given the set of Minkowski modes, one can obtain the Unruh modes $u^{\pm}_{\Omega,\Gamma}$ by a simple change of basis. Here $\Omega$ is the same label as for the Rindler modes and $\Gamma=R,L$ are extra indices. Positive and negative energy Minkowski modes do not mix and therefore the Unruh operators $C_{\Omega,\text{R}},C_{\Omega,\text{L}},D_{\Omega,\text{R}},D_{\Omega,\text{L}}$ annihilate the Minkowski vacuum as well. 
The Bogoliubov transformations between Unruh and Rindler operators takes the simple form 
\begin{align}\label{Unruhop}
\nonumber C_{\Omega,\text{\text{R}}}=&\left(\cosh r_{\Omega}\, c_{{\Omega},\text{I}}-\sinh r_{\Omega}\, d^\dagger_{{\Omega},\text{II}}\right),\\*
\nonumber C_{\Omega,\text{\text{L}}}=&\left(\cosh r_{\Omega}\, c_{{\Omega},\text{II}}-\sinh r_{\Omega}\, d^\dagger_{{\Omega},\text{I}}\right),\\*
\nonumber D^\dagger_{\Omega,\text{\text{R}}}=&\left(-\sinh r_{\Omega}\, c_{{\Omega},\text{I}}+\cosh r_{\Omega}\, d^\dagger_{{\Omega},\text{II}}\right),\\*
D^\dagger_{\Omega,\text{\text{L}}}=&\left(-\sinh r_{\Omega}\, c_{{\Omega},\text{II}}+\cosh r_{\Omega}\, d^\dagger_{{\Omega},\text{I}}\right),
\end{align}
where $\tanh r_{\Omega}=e^{-\pi \Omega}$.

The transformation between the Minkowski vacuum $\ket{0}_M$ and the Rindler vacuum $\ket{0}_R$ can be found in a standard way. 
We first introduce the generic Rindler Fock state $\ket{nm,pq}_{\Omega}$ as 
\begin{align}\label{shortnot}
\ket{nm,pq}_{\Omega}:=\frac{c_{{\Omega},\text{I}}^{\dagger n}}{\sqrt{n!}}\frac{d^{\dagger m}_{\Omega,\text{II}}}{\sqrt{m!}}\frac{d^{\dagger p}_{\Omega,\text{I}}}{\sqrt{p!}}\frac{c^{\dagger q}_{\Omega,\text{II}}}{\sqrt{q!}}\ket{0}_R
\end{align}
and the $\pm$ sign is again the notation for particle and antiparticle respectively. This allows us to write \cite{FABBRINAVARROSALAS}
\begin{equation}
\ket{0_\Omega}_\text{M}=\frac{1}{C^{2}}\sum_{n,m=0}^{+\infty}T^{n+m}\ket{nn,mm}_{\Omega},
\end{equation}
where $T:=\tanh r_\Omega$, $C:=\cosh r_\Omega$, $S:=\sinh r_\Omega$ and $\ket{0_\Omega}_\text{M}$ is a shortcut notation used to underline that each Unruh $\Omega$ is uniquely mapped to the corresponding Rindler $\Omega$.  

One particle Unruh states are defined as $\ket{1_{j}}^+_{\text{U}}=c_{\Omega,\text{U}}^\dagger\ket{0}_\text{M}$, $\ket{1_{j}}^-_{\text{U}}=d_{\Omega,\text{U}}^\dagger\ket{0}_\text{M}$
where the Unruh particle and antiparticle creation operator are defined as a linear combination of the two Unruh operators 
\begin{align}\label{creat}
c_{\Omega,\text{U}}^\dagger=&q_{\text{R}}C^\dagger_{\Omega,\text{R}}+q_{\text{L}}C^\dagger_{\Omega,\text{L}},\nonumber\\
d_{\Omega,\text{U}}^\dagger=&p_{\text{R}}D^\dagger_{\Omega,\text{R}}+p_{\text{L}}D^\dagger_{\Omega,\text{L}}.
\end{align}
$q_\text{\text{R}},q_\text{\text{L}},p_\text{\text{R}},p_\text{\text{L}}\in\mathbb{C}$ and satisfy $|q_\text{\text{R}}|^2+|q_\text{\text{L}}|^2=|p_\text{\text{R}}|^2+|p_\text{\text{L}}|^2=1$.
$p_\text{R,L}$  and $q_\text{R,L}$ are not independent. When restricted to the same Rindler wedge, for example the right wedge
\begin{align}
c_{\Omega,\text{U}}^\dagger\ket{0}_R=q_{\text{R}} C \ket{10,00}_{\Omega} \nonumber\\
d_{\Omega,\text{U}}^\dagger\ket{0}_{\text{R}}=p_{\text{L}} C \ket{00,10}_{\Omega}.
\end{align}
We require that in this case, and analogously when they are restricted to the left wedge, the Unruh particles and antiparticles have the \textit{same} interpretation of Rindler particle $\ket{10,00}_{\Omega}$ and antiparticle $\ket{00,10}_{\Omega}$. Therefore to be consistent with a particular choice of $q_\text{R}$ and $q_\text{L}$,  we must choose $p_\text{L}=q_\text{R}$ and $p_\text{R}=q_\text{L}$. \eqref{creat} reduces to
\begin{align}\label{creat}
c_{\Omega,\text{U}}^\dagger=&q_{\text{R}}C^\dagger_{\Omega,\text{R}}+q_{\text{L}}C^\dagger_{\Omega,\text{L}},\nonumber\\
d_{\Omega,\text{U}}^\dagger=&q_{\text{L}}D^\dagger_{\Omega,\text{R}}+q_{\text{R}}D^\dagger_{\Omega,\text{L}}.
\end{align}
Therefore, Unruh L and R excitations are given by
\begin{align}
\nonumber \ket{1_{k}}^+_\text{U}&=c_{k,\text{U}}^\dagger\ket{0}_U=q_{\text{R}}\ket{1_{\Omega,\text{R}}}^++q_{\text{L}}\ket{1_{\Omega,\text{L}}}^+\\
\ket{1_{k}}^-_\text{U}&=d_{k,\text{U}}^\dagger\ket{0}_U=q_{\text{L}}\ket{1_{\Omega,\text{R}}}^-+q_{\text{R}}\ket{1_{\Omega,\text{L}}}^-.\label{creat2}
\end{align}

\section{Particle and Anti-particle entanglement in non-inertial frames}\label{sec2}

Analyzing entangled states of Minkowski modes involves complicated computations since a single Minkowoski mode corresponds to an infinite superposition of Rindler modes. Therefore, for the sake of simplicity, we choose to analyze entangled states of Unruh modes instead \cite{Jorma}.

We have found the expressions for the vacuum and single Unruh and Rindler particle states. This  allows us to analyse the degradation of entanglement as seen by observers in uniform acceleration. Unruh modes with sharp frequency are delta-normalised. As discussed in \cite{Jorma}, one can always consider a superposition of Minkowski modes which will correspond to a distribution of Unruh frequencies $\Omega$. One can then choose the Minkowski distribution in such a way that the Unruh distribution will be peaked around some frequency $\Omega$.  Such Unruh wave packets would then be normalized. In the following we study the idealized case of Unruh modes with sharp frequencies. Although such states can be easily treated from a mathematical perspective, their physical interpretation requires deeper understanding. Preliminary results show that extended accelerated Unruh-deWitt detectors with a spatial Gaussian profile naturally couple to a peaked distribution of Unruh modes. These detectors will be used to analyze in more depth physical aspects of entangled states such as those considered in this work \cite{ANT}. 

We first consider the following family of maximally entangled states prepared by inertial observers Alice and Bob.
\begin{subequations}\label{states}
\begin{align}
\label{1e}\ket{\Psi_+}=&\frac{1}{\sqrt2}\left(\ket{0_\omega}_{\text M}\ket{0_\Omega}_\text{U}+\ket{1_\omega}^\sigma_{\text{M}}\ket{1_\Omega}^+_\text{U}\right)\\
\label{2e}\ket{\Psi_-}=&\frac{1}{\sqrt2}\left(\ket{0_\omega}_{\text M}\ket{0_\Omega}_\text{U}+\ket{1_\omega}^\sigma_{\text{M}}\ket{1_\Omega}^-_\text{U}\right)\\
\label{3e}\ket{\Psi_1}=&\frac{1}{\sqrt2}\left(\ket{1_\omega}^+_{\text M}\ket{1_\Omega}^-_\text{U}+\ket{1_\omega}^-_{\text{M}}\ket{1_\Omega}^+_\text{U}\right),
\end{align}
\end{subequations}
where $\text{U}$ labels bosonic Unruh modes and $\sigma=\pm$ denotes particle and antiparticle modes as usual. The entanglement will not depend on the value of Alice's Minkowski frequency~$\omega$, and we may regard the states \eqref{states} as parametrised by effectively just one parameter, the dimensionless Rindler frequency $\Omega$.

Rob does not naturally describe the states \eqref{states} with Minkowski coordinates but with Rindler coordinates. To take this into account we transform the Unruh modes to Rindler ones using \eqref{creat2}. After this transformation, the states become effectively a tri-partite system.  

Rob will study Rindler modes which have a fixed physical frequency $E$. Therefore, for any value of his proper acceleration $A$, he will analyze those states in \eqref{states} labeled by the corresponding dimensionless $\Omega=E/A$. In this sense, $\tanh r=\exp(-\pi E/A)$ and $A\rightarrow+\infty$ as $r\rightarrow+\infty$. Alternatively, Rob could have a fixed acceleration $A$ and would analyze Rindler modes with different frequencies $E$. He would then analyze those states in \eqref{states} labeled by the corresponding dimensionless $\Omega=E/A$.\\
As is commonplace in the literature, we define the Alice-Rob bi-partition as the Minkowski-region $I$ Rindler modes. 


To study distillable entanglement in this context we will employ the negativity $\mathcal{N}$, defined as the sum of the negative eigenvalues of the partial transpose density matrix \cite{negativity}. $\mathcal{N}\neq0$ is a sufficient condition for a state to be entangled. Two cases of interest will be considered. In the first case we assume that Alice and Rob have detectors which do not distinguish between particle and antiparticles. In this case, particles and antiparticles together are considered to be a subsystem. In the second case we consider that Rob has detectors which are only sensitive to particles or antiparticles and therefore antiparticle or particle states must be traced out. 

\subsection{Entanglement in states $\ket{\Psi_+}$ and $\ket{\Psi_-}$ }

\begin{figure}[!h]
\includegraphics[scale=0.9]{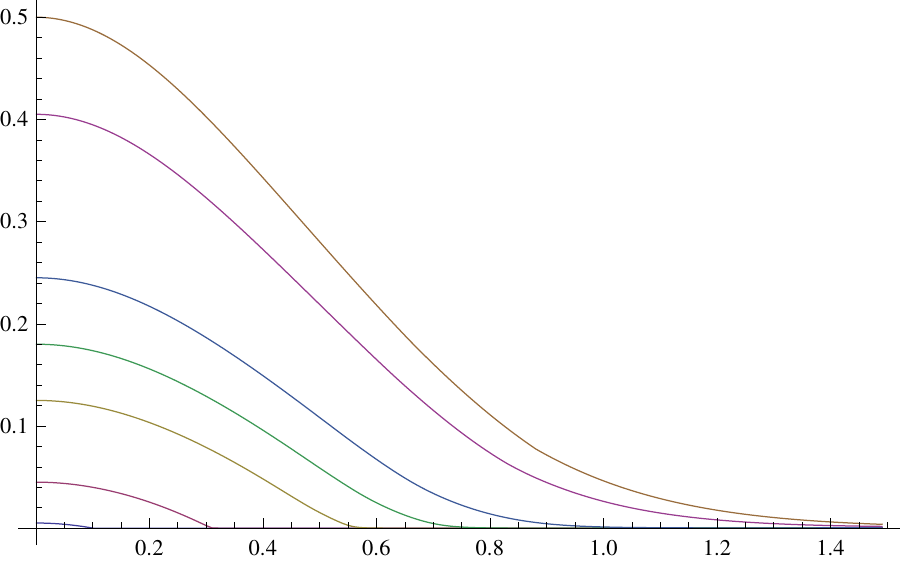}\caption{Negativity $\mathcal{N}$ as a function of $r$ for the state $\rho_{A-R}$. Curves are for $q_R=1,0.9,0.7,0.6,0.5,0.3,0.1$ from top to bottom.\label{FigureOne}}
\end{figure}

We start with states \eqref{1e} and \eqref{2e}.
To compute Alice-Rob partial density matrix  in \eqref{1e}  we trace over region II in $\proj{\Psi_+}{\Psi_+}$ and obtain,  
\begin{align}\label{ARd}
\rho_{A-R}^{PT}=&\frac{1}{2}\frac{1}{C^{4}}\sum_{n,m}T^{2n+2m}\left\{\right.\nonumber\\
&\left|0\right\rangle\left\langle 0\right|\otimes\left|n,m\right\rangle\left\langle n,m\right|\nonumber\\
+&\frac{1}{C^{2}}\left|1\right\rangle\left\langle 1\right|\otimes\left[(n+1)|q_{R}|^{2}\left| n+1,m\right\rangle\left\langle n+1,m\right|\right.\nonumber\\
+&T\sqrt{(n+1)(m+1)}q_{R}q_{L}^{*}\left|n+1,m+1\right\rangle\left\langle n,m\right|\nonumber\\
+&T\sqrt{(n+1)(m+1)}q_{L}q_{R}^{*}\left|n,m\right\rangle\left\langle n+1,m+1\right|\nonumber\\
+&\left. (m+1)|q_{L}|^{2}\left| n,m\right\rangle\left\langle n,m\right|\right]\nonumber\\
+&\frac{1}{C}\left|1\right\rangle\left\langle 0\right|\otimes\left[\sqrt{(n+1)}q_{R}^{*}\left|n,m\right\rangle\left\langle n+1,m\right|\right.\nonumber\\
+&\left.\left.T\sqrt{(m+1)}q_{L}^{*}\left|n,m+1\right\rangle\left\langle n,m\right|\right]+h.c.\right\}.
\end{align}
A major difference between the fermionic and the bosonic case is that in the latter, the Fock space is infinite dimensional in the particle number degree of freedom. In the present case it is therefore not possible to find the eigenvalues of the partial transpose density matrix analytically. However, we calculate $\mathcal{N}$ numerically and plot our results in Fig.\ref{FigureOne} as a function of $r=r(\Omega)$. 
We see that entanglement always vanishes in the infinite acceleration limit as for the uncharged bosonic case.

\begin{figure}[!h]
\includegraphics[scale=0.9]{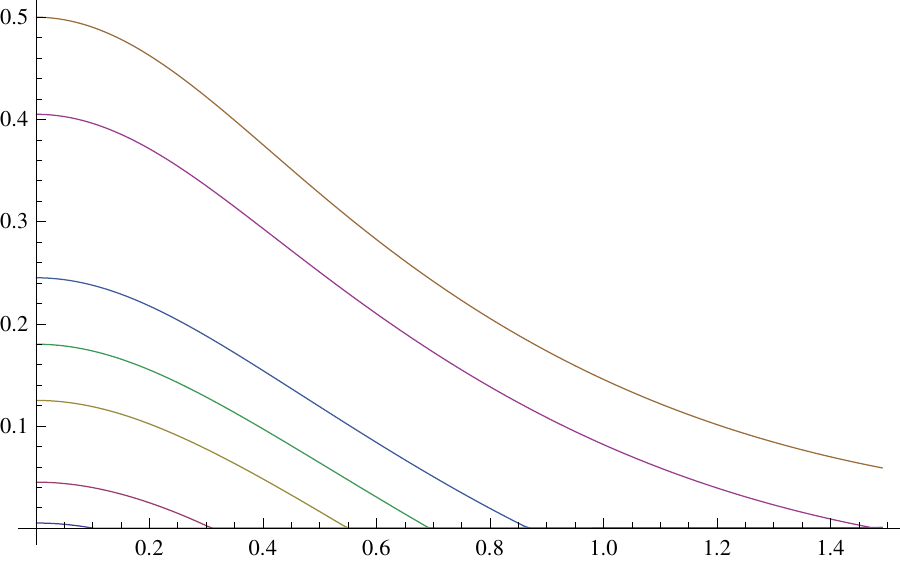}\caption{Negativity $\mathcal{N}$ as a function of $r$ for the state ${}_{-}\rho_{A-R}$. Curves are for $q_R=1,0.9,0.7,0.6,0.5,0.3,0.1$ from top to bottom.\label{FigureTwo}}
\end{figure}

We now analyse the entanglement when Rob is not able to detect antiparticles.  In this case Rob's particle modes are entangled with Alice's subsystem. Since Rob cannot detect antiparticles we must trace over all antiparticle states and therefore, \eqref{ARd}: ${}_{-}\rho_{A-R}^{PT}=\sum_{n}\bra{n_{\text{I}}^-}\rho^+_{AR}\ket{n_{\text{I}}^-}$. This yields
\begin{align}
{}_{-}\rho_{A-R}^{PT}=&\frac{1}{2}\sum_{n}\frac{T^{2n}}{C^2}\left\{
\left|0\right\rangle\left\langle 0\right|\otimes\left| n\right\rangle\left\langle n\right|+\right.\nonumber\\
+&\left|1\right\rangle\left\langle 1\right|\otimes\left[(n+1)\frac{1}{C^{2}}|q_{R}|^{2}\left| n+1\right\rangle\left\langle n+1\right|\right.\nonumber\\
+&\left. |q_{L}|^{2}\left| n\right\rangle\left\langle n\right| \right]\nonumber\\
+&\left.\left|1\right\rangle\left\langle 0\right|\otimes\left[\frac{1}{C}\sqrt{(n+1)}q_{R}^{*}\left| n\right\rangle\left\langle n+1\right|+h.c.\right]\right\}.\nonumber\\ \label{partially:transposed:reduced:state}
\end{align}
In this case we find analytical results. One can show that the partially transposed density matrix of the Alice-Rob bipartition has negative eigenvalues iff
\begin{align}
1\geq|q_{R}|^{2}>T^2\equiv|{}_{+}q_{R}^{AR}|^{2}.\label{cutoff}
\end{align}
This means that entanglement, quantified by $\mathcal{N}$, vanishes for finite acceleration. We plot the entanglement in this bipartition in Fig. \ref{FigureTwo}. 

The entanglement is always degraded in this bipartition and vanishes at finite $A$.
We will compare these results with those of the last part of section \ref{lastsection}. We stress that in the present case, the cutoff \eqref{cutoff} is the same for every eigenvalue of \eqref{partially:transposed:reduced:state}.

It is interesting to analyze the case where Rob and AntiRob's detectors are only sensitive to antiparticles. In this case one must trace over particle states. We obtain, \eqref{ARd}: ${}_{+}\rho_{A-R}^{PT}=\sum_{n}\bra{n}_{\text{I}}^+\rho^+_{AR}\ket{n}_{\text{I}}^+$, and therefore, 
\begin{align}
{}_{+}\rho_{A-R}^{PT}=&\frac{1}{2}\sum_{n}T^{2n}\left\{
\frac{1}{C^{2}}\left|0\right\rangle\left\langle 0\right|\otimes\left| n\right\rangle\left\langle n\right|\right.\nonumber\\
+&\left|1\right\rangle\left\langle 1\right|\otimes\left[(n+1)\frac{|q_{L}|^{2}}{C^{4}}+\left(T^{2}C^{2}\right)|q_{R}|^{2}\right]\left| n\right\rangle\left\langle n\right|\nonumber\\
+&\left.\left|1\right\rangle\left\langle 0\right|\otimes\left[\frac{T}{C^{3}}\sqrt{(n+1)}q_{L}^{*}\left| n+1\right\rangle\left\langle n\right|+h.c.\right]\right\}.\nonumber\\
\end{align}
In this case negative eigenvalues in the Alice-Rob partial transpose density matrix exist  iff
\begin{align}
|q_{L}|^{2}+T^{2}C^{2}|q_{R}|^{2}<0,
\end{align}
which can never be satisfied. Therefore, entanglement is always zero in this bipartition. This result is in clear contrast with the fermionic case in which entanglement is always created in this bipartition \cite{EDUIVY}. We therefore conclude that in the bosonic case the redistribution of entanglement between particles and antiparticles does not occur. 

The tensor product structure of the Hilbert space in the fermionic and the charged bosonic case plays an important role in the behavior of entanglement in the infinite acceleration limit.  In the case of neutral scalar fields there are no antiparticles and entanglement is completely degraded. One could expect that in the charged bosonic case transfer between particles and antiparticles might occur but we find that this is not the case. In the next section we will see more explicitly that the different statistics play a primary role in entanglement behavior.
We also notice that, as in \cite{EDUIVY}, these results have been computed for the initial state \eqref{1e}. One can easily find the result for the initial state \eqref{2e} by exchanging particle with antiparticle in all the previous calculations and conclusions.

\subsection{Entanglement in state $\ket{\Psi_1}$\label{lastsection}}

\begin{figure}[!h]
\includegraphics[scale=0.9]{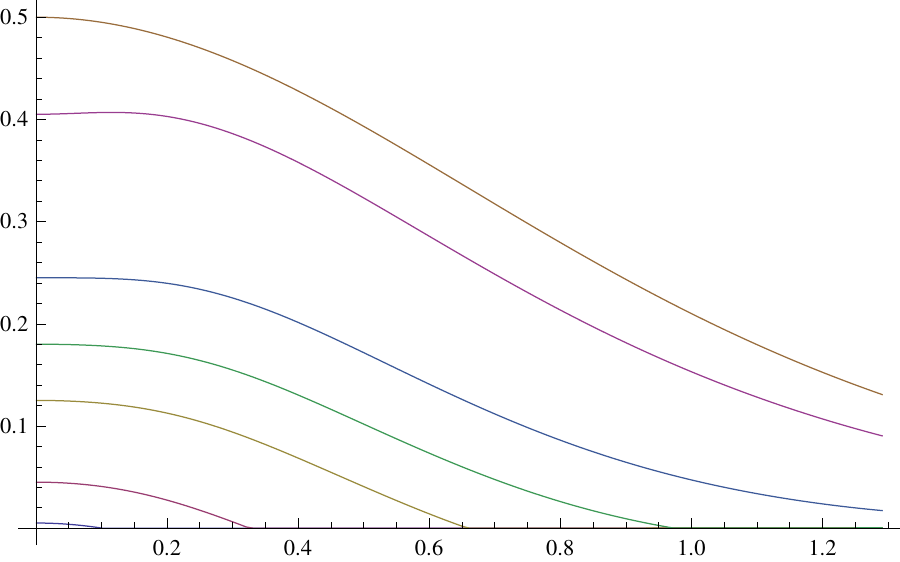}\caption{Negativity $\mathcal{N}$ as a function of $r$ for the state $\rho_{A-R}$. Curves are for $q_R=1,0.9,0.7,0.6,0.5,0.3,0.1$ from top to bottom.
\label{FigureThree}}
\end{figure}

We now study the entanglement in the state \eqref{3e}.
The density matrix for the subsystem Alice-Rob is obtained from $\proj{\Psi_1}{\Psi_1}$ by tracing over region $II$: 
\begin{align}
\rho_{A-R}^{PT}=&\frac{1}{2}\frac{1}{C^{6}}\sum_{n,m}T^{2n+2m}\left\{\right.\nonumber\\
&\left|-\right\rangle\left\langle -\right|\otimes\left[(n+1)|q_{R}|^{2}\left| n+1,m\right\rangle\left\langle n+1,m\right|\right.\nonumber\\
+&T\sqrt{(n+1)(m+1)}q_{R}q_{L}^{*}\left|n+1,m+1\right\rangle\left\langle n,m\right|\nonumber\\
+&T\sqrt{(n+1)(m+1)}q_{L}q_{R}^{*}\left|n,m\right\rangle\left\langle n+1,m+1\right|\nonumber\\
&\left. (m+1)|q_{L}|^{2}\left| n,m\right\rangle\left\langle n,m\right|\right]+\nonumber\\
+&\left|+\right\rangle\left\langle +\right|\otimes\left[(m+1)|q_{R}|^{2}\left| n,m+1\right\rangle\left\langle n,m+1\right|\right.\nonumber\\
+&T\sqrt{(n+1)(m+1)}q_{R}q_{L}^{*}\left|n+1,m+1\right\rangle\left\langle n,m\right|\nonumber\\
+&T\sqrt{(n+1)(m+1)}q_{L}q_{R}^{*}\left|n,m\right\rangle\left\langle n+1,m+1\right|\nonumber\\
&\left. (n+1)|q_{L}|^{2}\left| n,m\right\rangle\left\langle n,m\right|\right]\nonumber\\
+&\left|-\right\rangle\left\langle +\right|\otimes\left[T\sqrt{(m+1)(m+2)}q_{R}q_{L}^{*}\left| n,m+2\right\rangle\left\langle n,m\right|\right.\nonumber\\
+&\sqrt{(n+1)(m+1)}|q_{R}|^{2}\left|n,m+1\right\rangle\left\langle n+1,m\right|\nonumber\\
+&\sqrt{(n+1)(m+1)}T^2|q_{L}|^{2}\left|n,m+1\right\rangle\left\langle n+1,m\right|\nonumber\\
&\left.\left. T\sqrt{(n+1)(n+2)}q_{L}q_{R}^{*}\left| n,m\right\rangle\left\langle n+2,m\right|\right]+h.c.\right\}.\label{reduced:state:1}
\end{align}
As in the previous subsection, it is not possible to find an analytic expression for the eigenvalues of \eqref{reduced:state:1}. We calculate $\mathcal{N}$ numerically. We show our numerical results in Fig. \ref{FigureThree}

\begin{figure}[!h]
\includegraphics[scale=0.9]{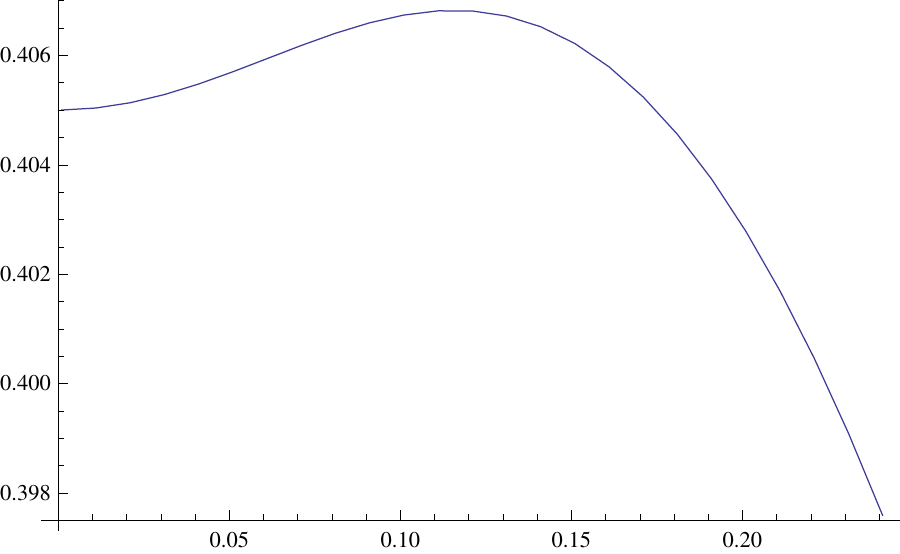}\caption{Negativity $\mathcal{N}$ as a function of $r$ for the state $\rho_{A-R}$ and $q_R=0.9$. A detail of the the curve shown in Fig. \ref{FigureThree} as the second curve from the top for $0<r<0.25$.
\label{FigureFive}}
\end{figure}

We find once more that entanglement is degraded in all cases  and vanishes in the limit of infinite acceleration.  Surprisingly, for small $r$, we notice that there is a range of values of $|q_R|$ where $\mathcal{N}$ is not a monotonically decreasing function of $r$. A similar behavior for a fermionic field was noticed in \cite{EduMille}. We show a sample in Fig. \ref{FigureFive}.

Assuming now that Rob's detector is only sensitive to particles, we trace over antiparticles in region $I$ and obtain
\begin{align}
{}_{-}\rho_{A-R}^{PT}=&\frac{1}{2}\sum_{n}T^{2n}\left\{\right.\nonumber\\
&\left|+\right\rangle\left\langle +\right|\otimes\left[\frac{|q_{R}|^{2}}{C^4}(n+1)\left| n+1\right\rangle\left\langle n+1\right|\right.\nonumber\\
+&\left.\frac{|q_{L}|^{2}}{C^2}\left| n\right\rangle\left\langle n\right|\right]\nonumber\\
+&\left|-\right\rangle\left\langle -\right|\otimes\left[\frac{|q_{L}|^{2}}{C^4}(n+1)
+\left.\frac{|q_{R}|^{2}}{C^2}\right]\left| n\right\rangle\left\langle n\right|\right.\nonumber\\
+&\left|+\right\rangle\left\langle -\right|\otimes\left[\frac{T}{C^{4}}\sqrt{(n+1)(n+2)}q_{L}q_{R}^{*}\left| n\right\rangle\left\langle n+2\right|\right.\nonumber\\
+&\left.\left.h.c.\right]\right\}. \label{Ku}
\end{align}

\begin{figure}[!h]
\includegraphics[scale=0.9]{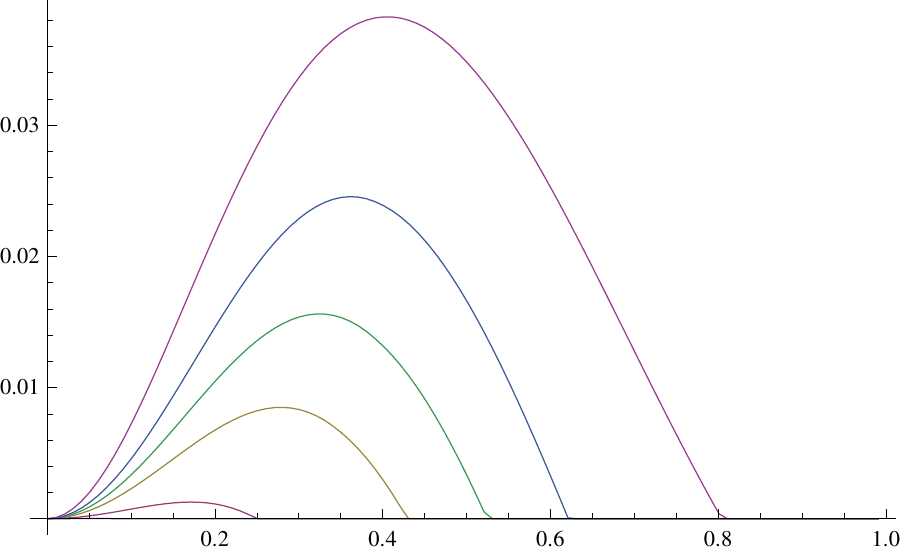}\caption{Negativity $\mathcal{N}$ as a function of  $r$ for the state $\rho_{A-R}$. Curves are for $q_R=1,0.9,0.7,0.6,0.5$ from top to bottom. \label{FigureFour}}
\end{figure}

We are able to analytically find the eigenvalues of the state \eqref{Ku}. Unlike the case for the state \eqref{partially:transposed:reduced:state}, where \textit{all} the eigenvalues could be negative if \eqref{cutoff} were satisfied, here we find that only a finite subset of the eigenvalues can be negative and such subset depends on $r$.
For this reason we compute $\mathcal{N}$ numerically. The entanglement for this scenario is plotted in Fig. \ref{FigureFour}. 

Assuming that Rob looks  only at antiparticles yields analogous results. 
We find that entanglement behaves very differently to the corresponding fermionic case where entanglement between Alice and Rob's particle (or antiparticle) sector is identically zero. However, here the entanglement grows with acceleration and reaches a maximum value after which it degrades. We trace the difference between the two cases down to extra terms of the form $\left| n\right\rangle\left\langle n+2\right|$ which appear in \eqref{Ku}. Clearly, no such fermionic Fock state as $\ket{(n+2)^{\pm}_{\Omega}}$ can exist due to Pauli exclusion principle.

\section{On the interpretation of field mode entanglement in non-inertial frames}\label{sec:middle}

Employing Unruh modes when analyzing the degradation of entanglement in non-inertial frames considerably simplifies mathematical computations since the Bogoliubov transformations between Unruh and Rindler modes are monochromatic \cite{Jorma}. However, we noticed that including Unruh modes requires a careful analysis of the results since the modes are parametrized by dimensionless $\Omega=E/A$. Therefore, our analysis involves a one parameter family of inertial orthogonal maximally entangled states. By fixing the Rindler frequency $E$ one analyzes the entanglement in a family of states, all of which share the same frequency $E$ as seen by observers with different proper acceleration $A$.  An alternative but also viable interpretation is the following: we analyze the entanglement of a family of states with different frequency $E$ as seen by the same observer moving with fixed proper acceleration $A$.

Furthermore, we have also pointed out that the physical interpretation of particle states which involve Unruh modes requires deeper understanding. These states are well defined mathematically, however their physical meaning is less understood (see for example, \cite{Tim}). Work in progress shows that a finite size Unruh-Dewitt detector in uniform acceleration naturally couples to peaked distributions of Unruh modes \cite{ANT}. Such detector model is employed to show that the single particle inertial states we consider here appear more mixed when probed by detectors with increasing acceleration. 

An alternative analysis on entanglement in non-inertial frames which does not employ Unruh modes but instead involves projective detectors \cite{evil1} is being considered in \cite{evil2}. 

\section{Conclusions}\label{conclusions}

Including antiparticles in the study of field mode entanglement in non-inertial frames has deepened our understanding of key features which explain the difference in behavior of entanglement  in the fermionic and bosonic case.  It was shown in \cite{EDUIVY} that in the fermionic case an entanglement redistribution between particle and antiparticle modes is responsible for the finite value of entanglement in the infinite acceleration limit. In particular, the relative redistribution for different particle and antiparticle bipartitions could be used to explain the behavior of the entanglement when particles and antiparticles were considered as a whole system. In this paper we included antiparticles in the study of bosonic entanglement by analyzing the charged bosonic case and computed the entanglement in the partitions that correspond to those considered for fermions in \cite{EDUIVY}. We showed that, due to the bosonic statistics,  there are substantial differences in the entanglement behavior when particles or antiparticles are not taken into account. We also found that there are values of the parameters for which the negativity is not a monotonically decreasing function of the acceleration. A similar behaviour for a fermionic field was observed in \cite{EduMille}. We confirmed that entanglement is always completely degraded in the infinite acceleration limit independently of the redistribution of entanglement between the particle and antiparticle bipartitions.

We have taken the opportunity to spell out in section \ref{sec:middle} an interpretational assertion that has been tacitly used in much of the recent literature on quantum correlations between inertial and uniformly accelerated observers. It would be an important question to identify classes of quantum observables for which this assertion would follow from interactions that are explicitly localized near Rob's word line.

\section*{Acknowledgements}
We would like to thank E. Martin-Martinez and  M.~Montero for interesting discussions, helpful comments and help in verifying the computations. We would like to thank A. Lee for help with Mathematica software.
I.~F. was supported by EPSRC [CAF Grant EP/G00496X/2]. A.~D. was supported in part by EPSRC [CAF Grant EP/G00496X/2] and by the Polish Ministry of Education award.
J.~L. was supported in part by STFC.

\bibliographystyle{apsrev}

\end{document}